\documentclass[prd,nofootinbib,preprint,superscriptaddress]{revtex4}
\usepackage{amsmath, amssymb, amsthm, graphicx, epsfig, fancyhdr,epsfig, slashed, mathrsfs}
\usepackage{braket}
\usepackage{tikzsymbols}
\usepackage{tikz}
\usepackage{tikz-feynman}
\tikzfeynmanset{compat=1.1.0}
\usepackage{natbib}
\usepackage{float}
\usetikzlibrary{shadows}
\usepackage{pifont}
\usepackage{thmtools}  
\usepackage{booktabs}
\usepackage{framed}
\usetikzlibrary{arrows.meta, positioning, shapes, decorations.markings}
\usepackage{caption}
\usepackage{adjustbox}

\theoremstyle{definition}

\theoremstyle{remark}

\usepackage{tikz,xcolor,hyperref}
\usepackage{subfig}

\definecolor{lime}{HTML}{A6CE39}
\DeclareRobustCommand{\orcidicon}{
	\begin{tikzpicture}
	\draw[lime, fill=lime] (0,0) 
	circle [radius=0.2] 
	node[white] {{\fontfamily{qag}\selectfont \tiny ID}};
	\draw[white, fill=white] (-0.0625,0.095) 
	circle [radius=0.007];
	\end{tikzpicture}
	\hspace{-2mm}
}

\foreach \x in {A, ..., Z}{\expandafter\xdef\csname orcid\x\endcsname{\noexpand\href{https://orcid.org/\csname orcidauthor\x\endcsname}
			{\noexpand\orcidicon}}
}

\newcommand{\be}{\begin{equation}}
\newcommand{\ee}{\end{equation}}
\newcommand{\bea}{\begin{eqnarray}}
\newcommand{\eea}{\end{eqnarray}}

\newcommand{\beq}{\begin{equation}}
\newcommand{\eeq}{\end{equation}}

\begin{document}

\title{Near-Resonant Thermal Leptogenesis}

\author{Angus Spalding \orcidC{}}
\email{angus.spalding1@gmail.com}
\affiliation{School of Physics and Astronomy, University of Southampton,
Southampton SO17 1BJ, United Kingdom}

\begin{abstract}
We study leptogenesis in the quasi-degenerate but non-resonant regime. Expanding the CP asymmetry parameter near degeneracy and imposing the conservative non-resonance condition that the mass splitting must be much greater than the right-handed neutrino decay rates $\Delta M > 100\Gamma_i$, yields the universal upper bound $\epsilon \leq 1/200$, independent of both the effective neutrino masses and the right-handed neutrino mass. We investigate vanilla and flavoured near-resonant leptogenesis and find that successful leptogenesis by right-handed neutrino decays can occur for $M \gtrsim 100~\mathrm{GeV}$ independent of washout regime, extending the viable parameter space of thermal leptogenesis down to the electroweak scale without invoking resonance. We also analyse near-resonant thermal leptogenesis during reheating and show that successful baryon asymmetry generation is compatible with reheating temperatures as low as $T_{RH}\simeq 10\rm GeV$ without relying on non-thermal production. Finally, we present a consistent framework for incorporating flavour effects in near-resonant leptogenesis during reheating. Overall, near-resonant thermal leptogenesis offers a controlled alternative regime to resonant leptogenesis, lowering the leptogenesis scale to the electroweak scale, without reliance on a disputed regulator used in resonant leptogenesis.

\end{abstract}

\maketitle

\tableofcontents
\section{Introduction}
The observation of neutrino oscillations \cite{Fukuda_2001, Fukuda_2002, Ahmad_2002, Ashie_2005, Abe_2008, Abe_2011, Abe_2016, Smirnov_2016, Aartsen_2018} confirmed that neutrinos have non-zero masses. This is inconsistent with the Standard Model, in which neutrinos are exactly massless, and therefore requires an extension of the standard model. One of the simplest and most well-established possibilities is the Type-I seesaw mechanism \cite{Minkowski:1977sc, Gell-Mann:1979vob, Yanagida:1979as, Mohapatra:1980yp}, which generates light neutrino masses through the introduction of heavy right-handed neutrinos. Through Yukawa interactions with the Higgs field, the new heavy states couple to the left-handed neutrinos; once the Higgs acquires its vacuum expectation value, these interactions generate light neutrino masses inversely proportional to the heavy mass scale. \\
The same heavy right-handed neutrinos responsible for generating light neutrino masses in the Type-I seesaw can also play a crucial role in explaining the observed baryon asymmetry of the Universe through leptogenesis \cite{Fukugita:1986hr}. In leptogenesis, the out-of-equilibrium and $CP$-violating decays of the lightest right-handed neutrino generate a net lepton asymmetry in the early Universe. This asymmetry is subsequently partially converted into a baryon asymmetry by electroweak sphalerons \cite{Luty:1992un, Giudice_2004, Covi_1996, Buchm_ller_2005, Sakharov:1967dj, Kolb:1979qa, Khlebnikov:1988sr}.\\
If the masses of the lightest two right-handed neutrinos are nearly degenerate the asymmetry generated is greatly enhanced. Maximal enhancement of the CP asymmetry is achieved in resonant leptogenesis \cite{Pilaftsis_1997, Anisimov_2006, Garbrecht_2014, Garny_2013, Riotto_2007, Klaric_2021, Granelli:2020ysj}, where the mass splitting between sterile neutrinos is comparable to their decay widths. In the fully resonant regime of leptogenesis, where the mass splitting between right-handed neutrinos is comparable to their decay widths, the CP asymmetry becomes formally divergent. Rendering this enhancement finite requires the introduction of an explicit regulating prescription. However, the form of this regulator is not unique and depends sensitively on the theoretical framework employed, with different analyses yielding quantitatively different results \cite{Pilaftsis_1997, Garbrecht_2014, Garny_2013, Riotto_2007, Klaric_2021}. This intrinsic prescription dependence renders the fully resonant regime theoretically delicate and highly model-dependent. This motivates us to focus on the near-resonant regime, which constitutes a physically distinct and theoretically controlled alternative to resonant leptogenesis, allowing sizeable CP violation without relying on regulator-dependent enhancements. As we will demonstrate, this scenario still considerably enlarges the parameter space for successful leptogenesis and can lower the required mass scale down to the electroweak scale within our setup.\\
Leptogenesis is most commonly studied in the context of thermal production of right-handed neutrinos, where these states are generated through scattering processes in the early Universe after reheating assuming radiation dominated universe throughout. However, in an inflationary cosmological history \cite{Guth:1980zm,Linde:1981mu,Lyth:1998xn}, a natural paradigm for leptogenesis to occur is during reheating \cite{Hahn_Woernle_2009, datta2023, datta2024, zhang2024, Haque:2023zhb, Haque:2024zdq, Chianese:2025mll}. We shall consider both scenarios in this work.\\
Leptogenesis is often analysed within the one-lepton or flavour-averaged approximation, in which all lepton flavours are treated as indistinguishable. In both thermal leptogenesis and leptogenesis during reheating, however, flavour effects can play a crucial role. Below temperatures of approximately \( 10^{12}\,\mathrm{GeV} \), the charged-lepton Yukawa interactions induce flavour decoherence, leading to distinct \( CP \)-violating effects in each flavour. A consistent treatment of these flavour dynamics is therefore essential for accurately predicting the final baryon asymmetry \cite{Nardi_2006, Nardi:2006fx, Abada_2006, Antusch_2006, Blanchet_2007, DeSimone:2006nrs, Cirigliano_2010, Simone_2007, Racker_2012, Moffat_2018, baker2024hotleptogenesis, Ulysses, Ulysses2, blanchet2013leptogenesisheavyneutrinoflavours, Granelli:2021fyc, Granelli:2025cho, Dev_2018, Dev_2014}. \\
Near-resonant leptogenesis has previously been investigated in an extended framework beyond the minimal setup \cite{datta2025gravitationalwavespectralshapes, ghoshal2025cosmicstringsgravitationalwave}. However, a corresponding analysis in the minimal leptogenesis scenario incorporating the new CP asymmetry parameter bound, including a consistent treatment of flavour effects, and during reheating has not yet been carried out. For clarity, we summarise in Table \ref{tab:whatsnew} which elements of our analysis are novel and which have been addressed in earlier work.\\
\begin{table}[h!]
\centering
\renewcommand{\arraystretch}{1.4}
\setlength{\tabcolsep}{12pt}
\begin{tabular}{|c|c|}
\hline
\textbf{Near-Resonant Leptogenesis Topic} & \textbf{Reference} \\
\hline
\textbf{CP Asymmetry Bound} & Derived in~\cite{datta2025gravitationalwavespectralshapes, ghoshal2025cosmicstringsgravitationalwave}. \\
\hline
\textbf{Vanilla Leptogenesis} & \textbf{This Work}. \\
\hline
\textbf{$\boldsymbol{U(1)_{B-L}}$-extended Leptogenesis} & Done in~\cite{datta2025gravitationalwavespectralshapes, ghoshal2025cosmicstringsgravitationalwave}. \\
\hline
\textbf{Leptogenesis during Reheating} & \textbf{This Work}. \\
\hline
\textbf{Flavoured Leptogenesis} & \textbf{This Work}. \\
\hline
\end{tabular}
\caption{\it Comparison of existing results and new contributions in near-resonant leptogenesis. Previous studies have derived the CP-asymmetry bound and explored near-resonant dynamics in an extended $U(1)_{B-L}$ framework, while this work presents the first analysis within the minimal vanilla leptogenesis scenario, including reheating dynamics and flavour effects.}
\label{tab:whatsnew}
\end{table}

The novelty of the present manuscript lies in analysing near-resonant leptogenesis within the minimal type-I seesaw and establishing its viability in the vanilla, flavoured, and reheating frameworks. In particular, we demonstrate that this quasi-degenerate but explicitly non-resonant regime successfully operates down to the electroweak scale.\\
\textit{This paper is organised as follows:} In Section \ref{sec:CP} we review the CP asymmetry parameter and derive the near-resonant bound. In Section~\ref{sec:vanilla}, we review the framework of thermal leptogenesis and study the parameter space for successful near-resonant thermal leptogenesis. We also study flavour effects in this regime. In Section~\ref{sec:reheat}, we study leptogenesis during reheating in a  universe initially dominated by a scalar field, the inflaton. This allows us to derive lower bound on the reheating temperature for successful leptogenesis. We then show how flavour effects can be incorporated into this framework. Finally, we summarise our findings in Section~\ref{sec:conclusion}.
\section{The Near-Resonant CP asymmetry: A Short Review}
\label{sec:CP}
In the type-I seesaw mechanism, three heavy right-handed Majorana neutrinos  $N_i$ are added to the Standard Model Lagrangian through the Yukawa and Majorana mass terms
\begin{equation}
    \mathcal{L} \supset 
    -\, Y_{\alpha i}\, \overline{L}_\alpha \, \tilde{H}\, N_i 
    - \frac{1}{2} M_i \, \overline{N_i^c} N_i 
    + \text{h.c.},
\end{equation}
where $L_\alpha$ denotes the lepton doublet of flavour $\alpha = e,\mu,\tau$, 
$\tilde{H} = i\sigma_2 H^*$ is the conjugate Higgs field, 
and $M_i$ are the heavy Majorana masses with $i$ being a flavour index.  
After electroweak symmetry breaking, the Higgs acquires a vacuum expectation value 
$\langle H \rangle = v / \sqrt{2}$, generating Dirac mass terms 
$m_D = v\, Y / \sqrt{2}$. In the limits $M_i\gg m_D$ the seesaw relation is achieved. We adopt the Casas–Ibarra parametrisation of the Yukawa matrix, which allows us to express the neutrino Yukawa couplings in terms of low-energy neutrino data, heavy neutrino masses, the PMNS matrix, and a complex orthogonal matrix.
\begin{equation}
    Y = \frac{1}{v}\, U\, \sqrt{m_\nu}\ R^T\ \sqrt{M}\ .
    \label{eq:CasasIbarra}
\end{equation}
A key quantity derived from this matrix, which controls both the decay rate of the right-handed neutrinos and the efficiency of leptogenesis, is the effective neutrino mass, defined for each heavy state as
\begin{equation}
    \tilde{m}_i = \frac{(Y^\dag Y)_{ii}v^2}{M_i}=\sum_{k} m_k\, |R_{ik}|^2,
    \label{eq:mtilde}
\end{equation}
which can be interpreted as a measure of the total coupling strength of the heavy neutrino $N_i$ to the thermal plasma. The decay width of $N_i$ at tree level is then
\begin{equation}
\Gamma_i(z_i)=\frac{M_i^2\,\tilde m_i}{8\pi v^2}\,\frac{K_1(z_i)}{K_2(z_i)} \, ,
\end{equation}
where $z_i=M_i/T$ and the ratio of modified Bessel functions accounts for the thermal averaging over the relativistic time dilation of decaying particles in the plasma.

\subsection{CP Asymmetry}
\begin{figure}[h!]
\centering
\begin{tikzpicture}[scale=1, baseline=(current bounding box.center)]

  \coordinate (Ni1) at (0,0);
  \coordinate (v1) at (1.5,0);
  \coordinate (ell1) at (3,0.8);
  \coordinate (phi1) at (3,-0.8);
  \draw[thick] (Ni1) -- (v1) node[midway, above] {\( N_i \)};
  \draw[thick] (v1) -- (ell1) node[midway, above ] {\( \ell \)};
  \draw[dashed] (v1) -- (phi1) node[midway, below ] {\( H \)};
  \node at (1.5, -1.2) {(a)};

  \begin{scope}[xshift=3.5cm]
    \coordinate (Ni2) at (0,0);
    \coordinate (loopL) at (1.2,0);
    \coordinate (loopR) at (2.4,0);
    \coordinate (v2) at (3.6,0);
    \coordinate (ell2) at (4.8,0.8);
    \coordinate (phi2) at (4.8,-0.8);
    \draw[thick] (Ni2) -- (loopL) node[midway, above] {\( N_i \)};
    \draw[thick] (loopR) -- (v2) node[midway, above] {\( N_j \)};
    \draw[thick] (v2) -- (ell2) node[midway, above ] {\( \ell \)};
    \draw[dashed] (v2) -- (phi2) node[midway, below ] {\( H \)};
    \draw[dashed] (loopL) arc[start angle=180, end angle=0, x radius=0.6cm, y radius=0.6cm] node[midway, above] {\( H \)};
    \draw[thick] (loopL) arc[start angle=-180, end angle=0, x radius=0.6cm, y radius=0.6cm] node[midway, below] {\( \ell \)};
    \node at (2.4, -1.2) {(b)};
  \end{scope}

  \begin{scope}[xshift=9cm]
    \coordinate (Ni3) at (0,0);
    \coordinate (vtx) at (1.2,0);
    \coordinate (Nj) at (2.4,0);
    \coordinate (ellOut) at (3.6,0.8);
    \coordinate (phiOut) at (3.6,-0.8);
    \coordinate (ellLoop) at (2.4,-0.8);
    \coordinate (phiLoop) at (2.4,0.8);
    \draw[thick] (Ni3) -- (vtx) node[midway, above] {\( N_i \)};
    \draw[thick] (vtx) -- (ellLoop) node[midway, below] {\( \ell \)};
    \draw[dashed] (vtx) -- (phiLoop) node[midway, above] {\( H \)};
    \draw[thick] (ellLoop) -- (phiLoop) node[midway, right] {\( N_j \)};
    \draw[thick] (phiLoop) -- (ellOut) node[midway, above right] {\( \ell \)};
    \draw[dashed] (ellLoop) -- (phiOut) node[midway, below right] {\( H \)};
    \node at (1.8, -1.2) {(c)};
  \end{scope}

\end{tikzpicture}
\caption{\it Feynman diagrams contributing to the CP asymmetry: (a) tree-level, (b) self-energy, and (c) vertex diagrams.}
\label{fig:cpasymmetrydiagrams}
\end{figure}
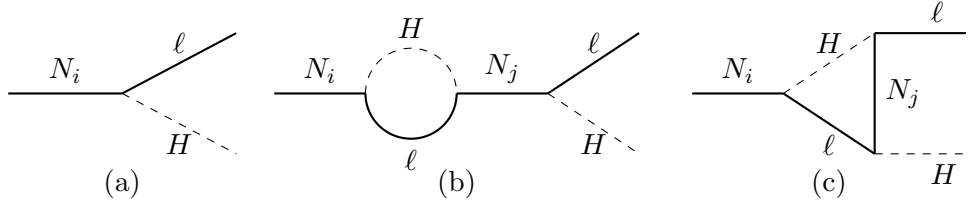

In leptogenesis, a lepton asymmetry is generated through the $CP$-violating decays of heavy right-handed neutrinos into a Higgs boson and a lepton at one loop order. The relevant decay processes are depicted in Figure \ref{fig:cpasymmetrydiagrams}.
The interference between these amplitudes generates different decay rates for $N_i \to L H$ and $N_i \to \Bar{L} H^\dagger$, giving rise to a net lepton asymmetry. The amount of CP violation is measured by the $CP$ asymmetry parameter defined in terms of decay rates
\begin{equation}
    \epsilon_i = 
    \frac{\Gamma(N_i \to L H) - \Gamma(N_i \to \Bar{L} H^\dagger)}
         {\Gamma(N_i \to L H) + \Gamma(N_i \to \Bar{L} H^\dagger)}\ .
\end{equation}
At one loop, this can be expressed in terms of right-handed neutrino masses and the yukawa couplings ~\cite{Covi_1996, Buchm_ller_2005, Giudice_2004, Di_Bari_2012}
\begin{equation}
    \epsilon_i = 
    \frac{1}{8\pi}
    \sum_{j \neq i}
    \frac{{\rm Im}[(Y^\dagger Y)_{ij}^2]}{(Y^\dagger Y)_{ii}}\,
    f\!\left(\frac{M_j^2}{M_i^2}\right),
    \label{eq:eps_general}
\end{equation}
where the loop function encodes the mass dependence on the CP violation parameter
\begin{equation}
    f(x) = \sqrt{x}\left[(1 + x)\ln\!\left(\frac{1 + x}{x}\right) - \frac{2 - x}{1 - x}\right]\ .
\end{equation}
From this expression, we observe that $\epsilon$ becomes large when the masses of two right-handed neutrinos approach degeneracy, $M_i \approx M_j$, in extreme degeneracy where the mass splitting is of the order of the right-handed neutrino decay rates this leads to the phenomenon of resonant leptogenesis \cite{Pilaftsis_2004}. This scenario requires a refined treatment of the self-energy contribution, which dominates in this limit and develops a regulated enhancement. Various prescriptions for regulating the divergent behaviour have been proposed in the literature \cite{Klaric_2021, Pilaftsis_2004, Riotto_2007, Garbrecht_2014, Garny_2013, Anisimov_2006}. Owing to this ambiguity, in the present study, we refrain from analysing this regime. To this end, we shall impose the condition that the mass splitting is much greater than the decay rates of the right-handed neutrinos \cite{Moffat_2018}. 

\subsection{Non-Resonant bound on the CP asymmetry}
In this subsection we derive the near-resonant bound on the CP asymmetry as derived in \cite{ghoshal2025cosmicstringsgravitationalwave, datta2025gravitationalwavespectralshapes}. In the quasi-degenerate regime where $M_i \simeq M_j$, we can expand the form of the CP asymmetry parameter in \ref{eq:eps_general} in small $\Delta M/M$ gives
\begin{equation}
\epsilon_i \;\simeq\; \frac{1}{16\pi}\,\frac{M_i}{\Delta M}\,
\frac{Im\big[(y^\dagger y)_{ij}^2\big]}{(y^\dagger y)_{ii}},\ \leq \frac{1}{16\pi}\,\frac{M_i}{\Delta M}\,(y^\dagger y)_{jj}.
\end{equation}
where in the last step we have bounded the numerator with the Cauchy-Schwartz inequality. We remain in the non-resonant regime to avoid the need for a regulator and the associated theoretical ambiguities of the resonant case, where the CP asymmetry is highly sensitive to the mass splitting and the chosen regularisation scheme. To remain in the non-resonant regime, the mass splitting must greatly exceed both decay widths \cite{Pilaftsis_2004, Moffat_2018}, we take this to be a conservative one hundred times the decay widths following \cite{Moffat_2018},
\begin{equation}
\Delta M > 100\,\Gamma_1,
\quad
\Delta M > 100\,\Gamma_2, \quad \Gamma_i = \frac{(y^\dagger y)_{ii}}{8\pi}\,M_i\ .
\end{equation}
Evaluating $\epsilon_1$, the condition from $\Gamma_1$ yields an upper limit on the CP-violating parameter, expressed in terms of the ratio of effective neutrino masses,
\begin{equation}
|\epsilon_1| \;\lesssim\; \frac{1}{200}\,\frac{(y^\dagger y)_{22}}{(y^\dagger y)_{11}}\,\frac{M_2}{M_1}\simeq\frac{1}{200}\frac{\tilde m_2}{\tilde m_1}\ .
\end{equation}
In contrast, the constraint from $\Gamma_2$ removes any dependence on masses or Yukawas,
\begin{equation}
|\epsilon_1| \;\lesssim\; \frac{1}{200}.
\end{equation}
Both conditions must hold meaning the corresponding limits for decays of the heavier right-handed neutrino are
\begin{equation}
|\epsilon_2| \;\leq\; \min\!\left[\frac{1}{200},
\;\frac{1}{200}\,\frac{\tilde m_1}{\tilde m_2}\right].
\end{equation}
These results are summarised in Table \ref{tab:epsbounds}. Since both non-resonance conditions must hold simultaneously, the true universal ceiling is independent of both the right-handed neutrino mass and the corresponding effective neutrino mass.
\begin{equation}
|\epsilon_i| \;\leq\; \frac{1}{200}\ . 
\end{equation}
Thus, outside the resonant regime, the CP asymmetry per decay cannot exceed the percent level, independently of right-handed neutrino masses and Yukawa couplings. If one chooses to impose the resonance condition $\Delta M > a \Gamma_i$ more or less strictly, with $a=100$ our baseline, the corresponding bound on the CP asymmetry follows directly as $\epsilon<1/2a$.

\begin{table}[h!]
\centering
\renewcommand{\arraystretch}{1.4}
\setlength{\tabcolsep}{10pt}
\begin{tabular}{|c|c|c|}
\hline
Hierarchy & Bound on $\epsilon_1$ & Bound on $\epsilon_2$ \\
\hline
$\tilde m_1 < \tilde m_2$ 
& $\displaystyle |\epsilon_1| \leq \frac{1}{200}$ 
& $\displaystyle |\epsilon_2| \leq \frac{1}{200}\,\frac{\tilde m_1}{\tilde m_2}$ \\
\hline
$\tilde m_2 < \tilde m_1$ 
& $\displaystyle |\epsilon_1| \leq \frac{1}{200}\,\frac{\tilde m_2}{\tilde m_1}$ 
& $\displaystyle |\epsilon_2| \leq \frac{1}{200}$ \\
\hline
\end{tabular}

\captionsetup{justification=raggedright, singlelinecheck=false, font=it}
\caption{Analytic bounds on the CP asymmetries, $\epsilon_{1,2}$, in the quasi-degenerate regime imposing the non-resonant condition $\Delta M > 100 \Gamma_i$.}
\label{tab:epsbounds}
\end{table}
We now apply this bound on the CP asymmetry to vanilla thermal leptogenesis and leptogenesis during reheating to find the parameter space for successful leptogenesis.
\section{Thermal Near-Resonant Leptogenesis}
\label{sec:vanilla}
The origin of the matter–antimatter asymmetry remains one of the central open problems in cosmology and particle physics. The baryon asymmetry of the Universe has been precisely determined through independent observations of the cosmic microwave background~\cite{Planck2018} and Big Bang nucleosynthesis~\cite{ParticleDataGroup:2020ssz}
\begin{equation}
    Y_B=\frac{n_B}{s}= 8.87\times 10^{-11}\ .
    \label{eq:BAU}
\end{equation}
Thermal leptogenesis provides the most minimal and widely studied framework for generating the baryon asymmetry of the Universe, relying solely on the out-of-equilibrium decays of heavy right-handed neutrinos produced through standard thermal processes. In this section, we systematically study this paradigm in the near-resonant regime within the vanilla treatment, before subsequently incorporating flavour effects into the analysis.
\subsection{Boltzmann Equations}
To compute the baryon asymmetry, we solve the Boltzmann equations governing the evolution of the lepton asymmetry and the relevant particle abundances. These are very well established equations tracking the abundances of two decaying right-handed neutrinos and the lepton-anti lepton asymmetry. The lepton asymmetry equation contains two distinct contributions: a source term, which generates the asymmetry through $CP$-violating decays, and a washout term, which accounts for inverse decays and $\Delta L = 2$ scattering processes \cite{Fong_2012, Ulysses, Ulysses2}. 
\begin{equation}
\begin{aligned}
    \frac{dY_{N_i}}{dz}&=-D_i(z)\Big(Y_{N_i}-Y_{N_i}^{\rm eq}(z)\Big) \\
    \frac{dY_{B-L}}{dz}& = \sum_i \epsilon_i D_i(z) \Big(Y_{N_i} - Y_{N_i}^\text{eq}(z)\Big) - W_i(z) Y_{B-L}\ .
\end{aligned}
\end{equation}
where $z=M/T$ is the standard dimensionless variable parametrising the evolution, $D$ ($W$) denotes the decay (washout) and are standard results in the literature \cite{Moffat_2018} that depend on the right handed neutrino mass, the associated effective neutrino mass, and temperature.  
\begin{equation}
D_i(z) = K_i \, z \, \frac{K_1(z)}{K_2(z)} , \quad W_i(z) = \frac{1}{4} K_i \, K_1(z) \, z^3
\end{equation}
with $K_1$ and $K_2$ the modified Bessel functions of the first and second kind, and
\begin{equation}
K \equiv \frac{\tilde{\Gamma}}{H(T=M_i)}, 
\qquad 
\tilde{\Gamma}_i = \frac{M_i \,(Y^\dagger Y)_{ii}}{8\pi} .
\label{eq:Ki}
\end{equation}
Here $H$ is the Hubble expansion rate assuming radiation domination. The $Y_i^{\rm eq}$ terms are the equilibrium abundances of the ith right-handed neutrino,
\begin{equation}
Y_i^{\mathrm{eq}} = \frac{45}{4\pi^4} \, \frac{g_i}{g_*} \, z_i^2 \, K_2\!\left(z_i\right),
\end{equation}
The final baryon asymmetry is then proportional to $Y_{B-L}$ through the sphaleron conversion factor
\begin{equation}
    Y_B=\frac{28}{79}\ Y_{B-L}\ .
\end{equation}
The vanilla treatment considers only a single effective lepton flavour. Flavour effects~\cite{Nardi:2006fx,Abada_2006,Antusch_2006,Blanchet_2007,DeSimone:2006nrs,Cirigliano_2010,Simone_2007,Racker_2012} become relevant for non-negligible washout and 
\( M_1 \lesssim 10^{12}\,\mathrm{GeV} \), 
a mass range that is of interest in the present work and will be discussed in detail in Section~\ref{sec:flavour}. Since the inclusion of flavour effects can only weaken the resulting bounds, we first focus on the vanilla framework and show that successful leptogenesis can already be achieved down to the electroweak scale within this minimal treatment.

\subsection{Results}
\begin{figure}[H]
    \centering
    \includegraphics[width=\textwidth]{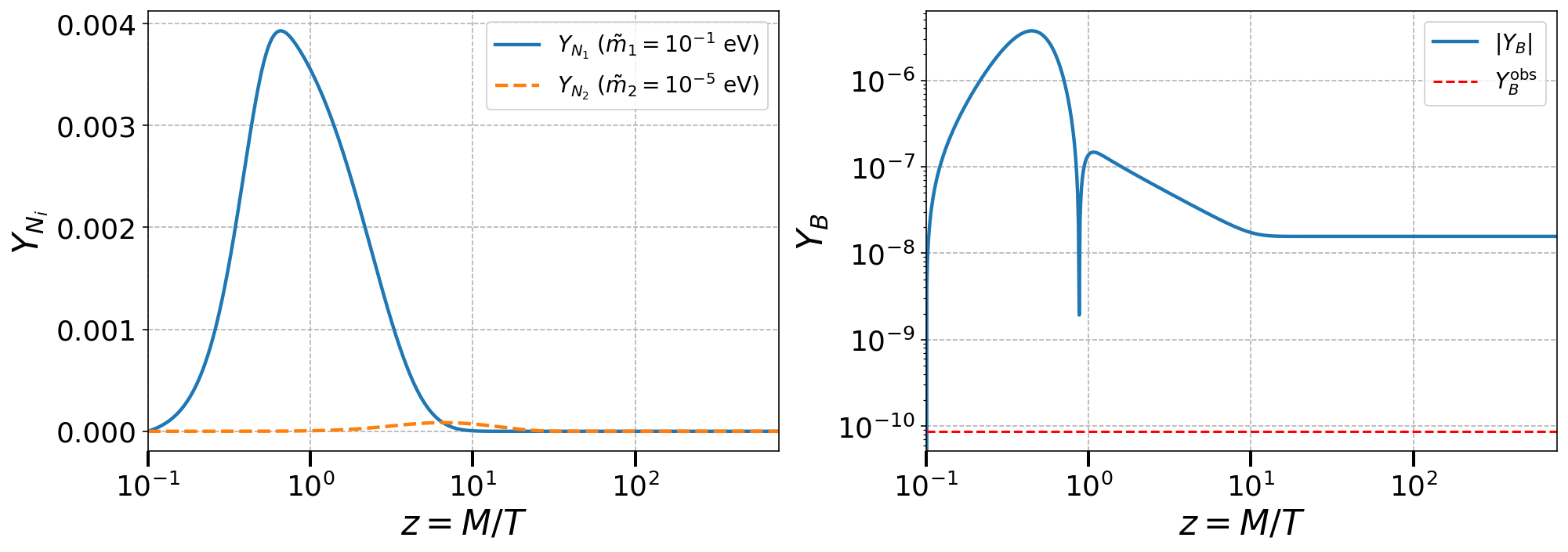}
    \caption{\it Benchmark result for completed leptogenesis taking the maximum bound on the CP asymmetry with parameters $M = 10^{5}\,\text{GeV}$, $\tilde m_1 = 10^{-1}$ eV and $\tilde m_2= 10^{-5}\rm\ eV$. \textbf{Left Panel:} The evolution of the right-handed neutrino abundances. The right-handed neutrino with the larger corresponding $\tilde m$ is able to be produced at greater abundance before becoming non-relativistic and decaying. \textbf{Right Panel:} the evolution of the baryon asymmetry. The resulting baryon asymmetry exceeds the observed value by several orders of magnitude, illustrating the high efficiency of baryon asymmetry generation in near-resonant leptogenesis. }
    \label{fig:benchmark}
\end{figure}

\begin{figure}[H]
    \centering
    \includegraphics[width=\textwidth]{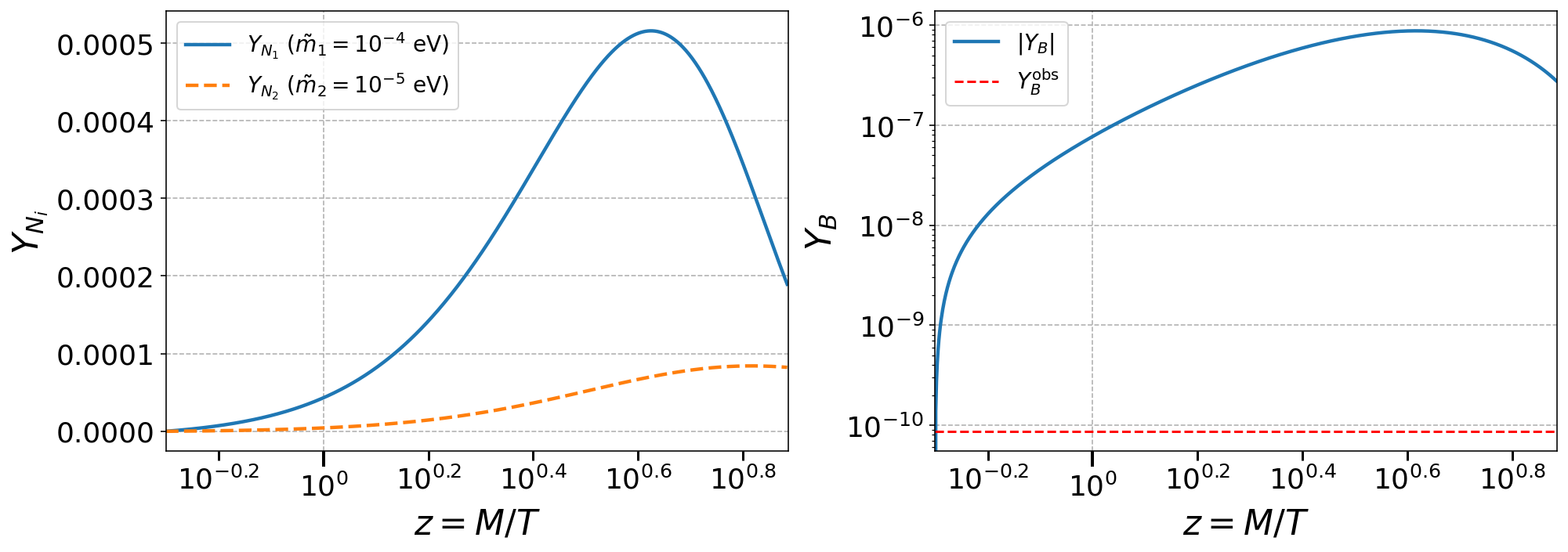}
    \caption{\it Benchmark result for low scale near-resonant leptogenesis with parameters are $M = 10^{3}\,\text{GeV}$, $\tilde m_1 = 10^{-4}$ eV and $\tilde m_2= 10^{-5}\rm\ eV$ and shows the evolution of the bolztmann equations up to electroweak symmetry breaking. \textbf{Left Panel:} the evolution of the right-handed neutrino abundances, illustrating that their production and decay remain suppressed due to the smaller effective neutrino masses and that the decays are not complete by electroweak symmetry breaking. \textbf{Right Panel:} the baryon asymmetry evolution and demonstrates that despite the incomplete decays, the CP asymmetry is sufficiently enhanced that the observed baryon asymmetry is still reproduced.}
    \label{fig:benchmark2}
\end{figure}

\begin{figure}[H]
    \centering
    \includegraphics[width=0.48\textwidth]{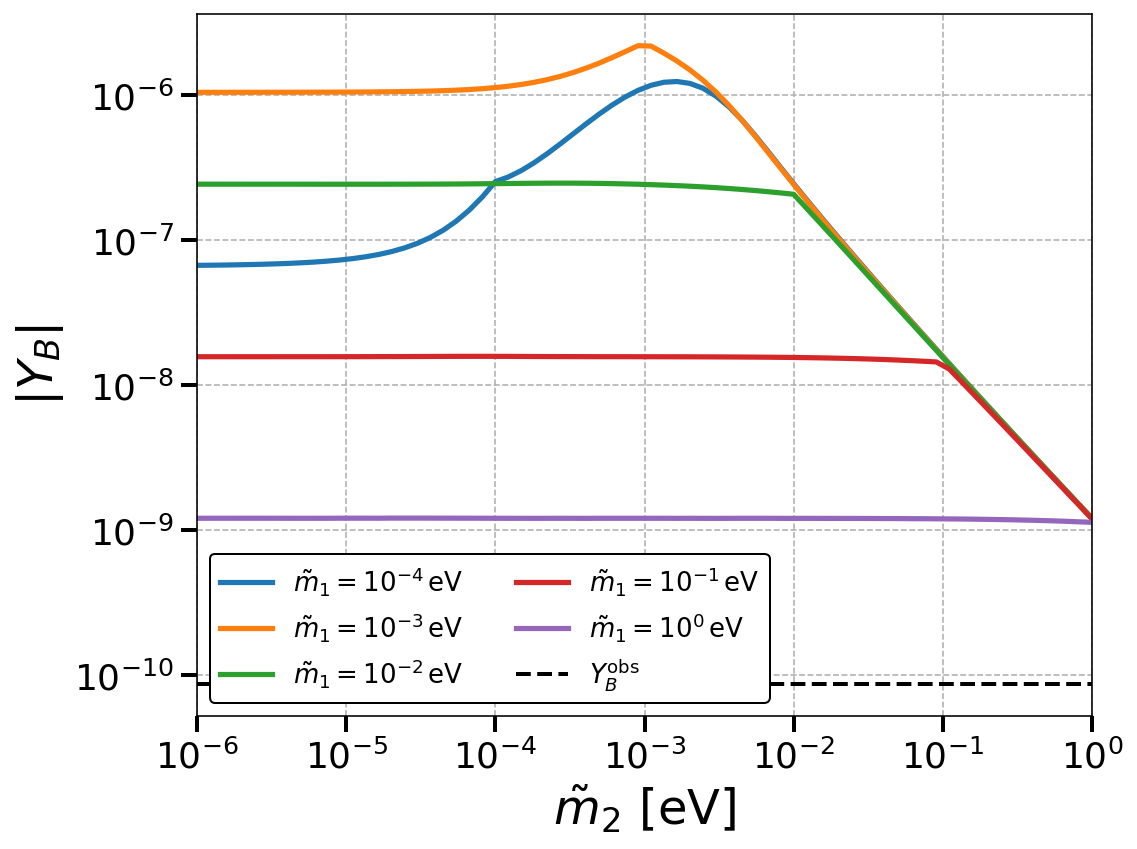}
    \includegraphics[width=0.48\textwidth]{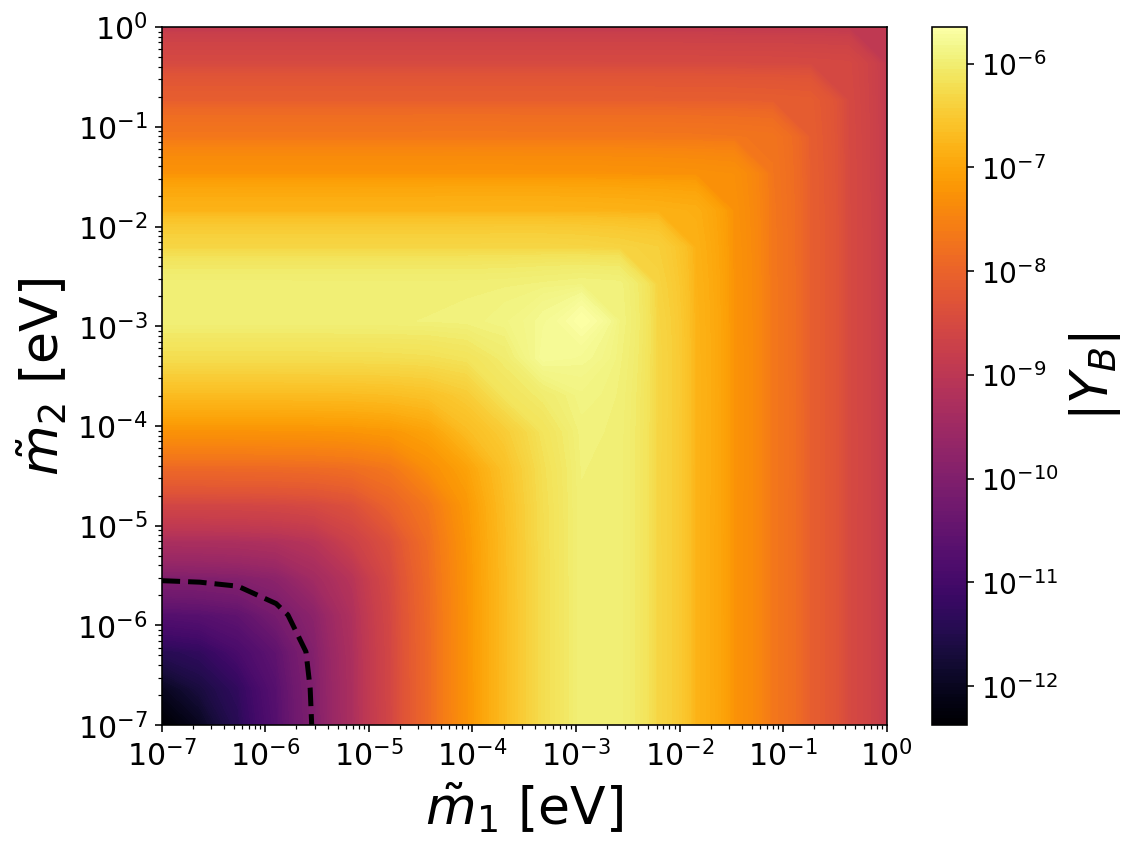}
    \caption{\it Maximum baryon asymmetry from scans over effective neutrino masses for $M = 10^{5}\,\text{GeV}$. \textbf{Left Panel:} we fix $\tilde m_1$ and scan over $\tilde m_2$. We can see that for $\tilde m_2<<\tilde m_1$ the result is independent of $\tilde m_2$ and for $\tilde m_2>>\tilde m_1$ the result is independent of $\tilde m_1$. We can thus conclude that if we have a hierarchy in effective neutrino masses the baryon asymmetry is determined by the larger effective neutrino mass. \textbf{Right Panel:} shows the full scan demonstrating that if either effective neutrino mass is greater than $\simeq 3\times 10^{-6}$ eV then the observed baryon asymmetry can be reproduced.}
    \label{fig:thermalscan1}
\end{figure}

\begin{figure}[H]
    \centering
    \includegraphics[width=\textwidth]{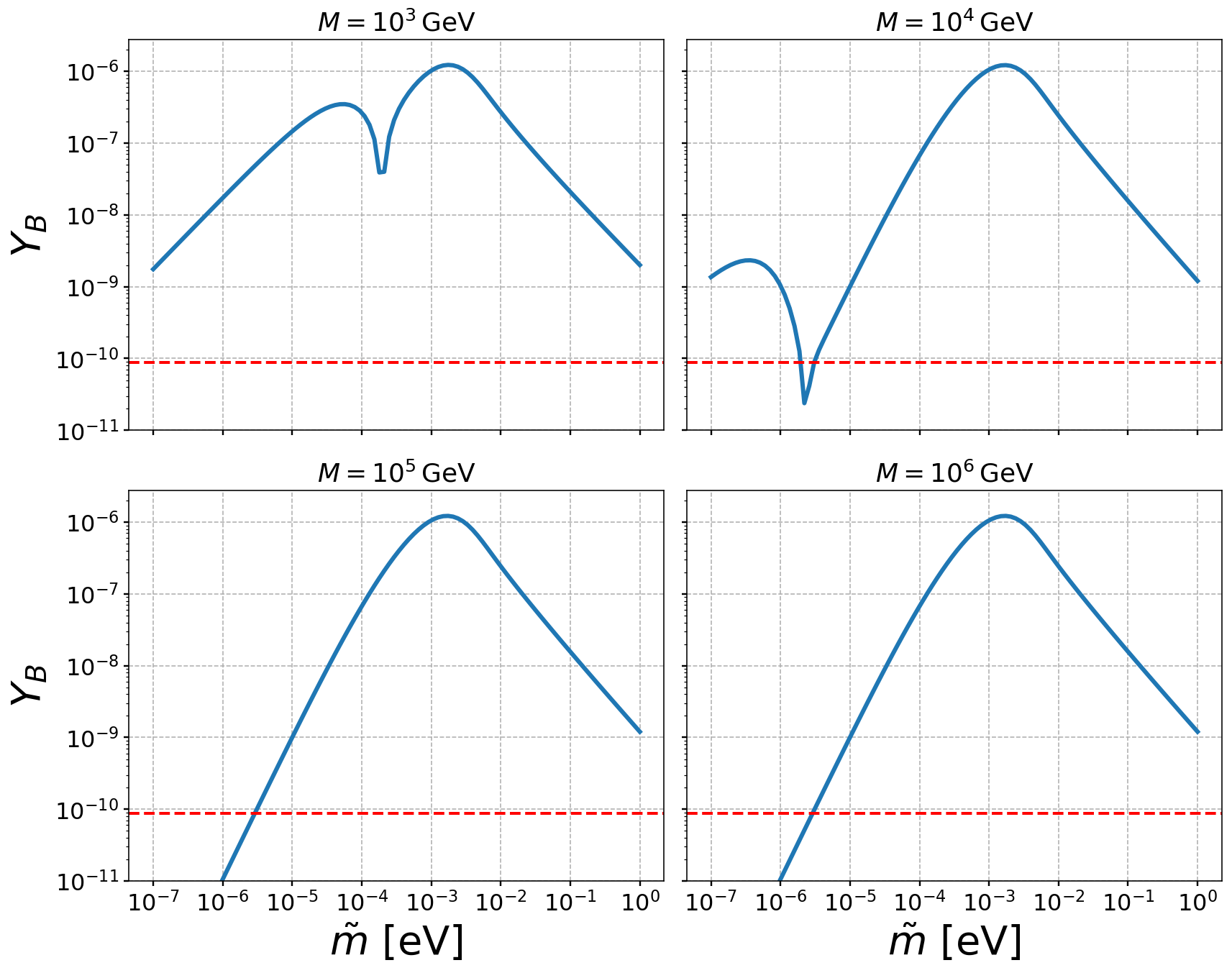}
    \caption{\it The maximum baryon asymmetry for various right-handed neutrino masses with the largest effective neutrino mass scanned over for zero initial conditions. The horizontal dashed line indicates the observed baryon asymmetry. The curves exhibit the characteristic suppression at small $\tilde m$ due to inefficient right-handed neutrino production and at large $\tilde m$ due to strong washout. For lower $M$, additional structure appears when electroweak symmetry breaking and sphaleron freeze-out occur during the intrinsic sign change in the baryon-asymmetry evolution.
}
    \label{fig:thermalscan3}
\end{figure}

As near-resonant leptogenesis allows the scale of baryogenesis to be significantly lowered, care must be taken to evaluate the generated asymmetry only up to the electroweak symmetry breaking temperature, \(T_{\rm EW} = 130\,\mathrm{GeV}\). In this regime, the right-handed neutrinos may not have fully decayed by electroweak symmetry breaking, while still generating a sufficient baryon asymmetry. We assume that a period of inflation occurred prior to leptogenesis, such that the initial abundances of both right-handed neutrinos and any pre-existing asymmetry are negligible. A fully completed leptogenesis benchmark, in which the right-handed neutrinos decay entirely before electroweak symmetry breaking, is shown in figure \ref{fig:benchmark}.
The right-handed neutrino associated with the larger effective neutrino mass \(\tilde m\) reaches a higher abundance before becoming non-relativistic, as inverse decays are more efficient in this regime. The feature appearing around \(z \simeq 1\) corresponds to a change in the sign of the generated baryon asymmetry. The resulting baryon asymmetry exceeds the observed value by several orders of magnitude, illustrating the high efficiency of baryon asymmetry generation in near-resonant leptogenesis. Since the upper bound on the CP asymmetry \(\epsilon\) is independent of the right-handed neutrino mass \(M\), and all remaining quantities entering the Boltzmann equations depend only on the dimensionless variable \(z = M/T\), the evolution shown here is identical for all values of \(M\).
This plot will be identical for all masses of the right-handed neutrino that ensure decays have completed before electroweak symmetry breaking. If a different choice of the non-resonance parameter were adopted, defined through the condition $\Delta M > a\,\Gamma_i$, where we take the conservative benchmark $a=100$, the maximal CP asymmetry would scale as $1/a$, and the resulting baryon asymmetry would scale proportionally. For example, taking $a=50$ would increase the maximal asymmetry by a factor of two, while $a=200$ would reduce it by a factor of two. This scaling therefore applies uniformly to all subsequent baryon asymmetry plots shown in this paper. For low scale right-handed neutrino masses and small effective neutrino masses leptogenesis isn't completed but instead the electroweak phase transition halts it and fixes the baryon asymmetry as sphaleron processes freeze out. A benchmark showing this is shown in Figure \ref{fig:benchmark2}.
This shows that for low-scale near resonant leptogenesis decays haven't completed yet before electroweak symmetry breaking however the generation of the asymmetry is so large that the observed baryon asymmetry is still easily reproduced. Generally the form of the baryon asymmetry is such that by $z\lesssim 1$ indicating that CP-violating decays can successfully realise baryogenesis even for mass scales slightly below the electroweak phase transition $M\gtrsim 100$ GeV. Below this mass range enters the regime of ARS leptogenesis where leptogenesis is driven oscillations rather than decays \cite{Drewes_2018, Akhmedov_1998, Klaric_2021}. As discussed above, the non-resonance condition introduces a quantitative parameter $a$, which we fix to $a=100$ as a conservative benchmark in this work. Across much of the viable parameter space, however, the maximal baryon asymmetry exceeds the observed value by several orders of magnitude. Consequently, moderate variations in $a$ do not qualitatively modify the viable mass window, nor the conclusion that successful leptogenesis can be realised down to the electroweak scale. We now look to find the specific values of the effective neutrino masses that will allow for successful leptogenesis. We begin by identifying which right-handed neutrinos play the dominant role in determining the baryon asymmetry. To this end in figure \ref{fig:thermalscan1} we plot the maximum baryon asymmetry achieved scanning over effective neutrino mass values.
We observe that in the regime $\tilde m_2 \ll \tilde m_1$ the generated baryon asymmetry becomes independent of $\tilde m_2$, while for $\tilde m_2 \gg \tilde m_1$ it is instead independent of $\tilde m_1$. This demonstrates that, in the presence of a hierarchy among the effective neutrino masses, the baryon asymmetry is controlled by the larger effective neutrino mass. Physically, this occurs because the upper bound on the CP asymmetry is maximised for the larger effective neutrino mass, and because a larger $\tilde m$ enhances the thermal production of right-handed neutrinos from the plasma. Even though for very large values of $\tilde m_2$ the washout is very strong the effects of the increase right-handed neutrino production and CP violation dominates the competing effects.
In the right panel we present a systematic scan over the effective neutrino mass parameters and that for effective neutrino masses above $\simeq 3\times 10^{-6}$ eV the observed baryon asymmetry is readily reproduced. For small values of $\tilde m$ very few decays or inverse decays occur and as such a smaller asymmetry is generated. At the other extreme for very large $\tilde m$ the washout term removes this asymmetry. To now see the effect of the various masses we will assume a hierarchy in effective neutrino masses We fix $\tilde m_2=10^{-10}$ eV, a negligible value, and scan over values of $\tilde m_1$ the result is shown in figure \ref{fig:thermalscan3} for various right-handed neutrino masses. 
The general features of the figure can be understood from the competing effects of right-handed neutrino production and washout. For very small effective neutrino masses, the production of right-handed neutrinos is inefficient, leading to a suppressed baryon asymmetry. In addition, due to our choice of vanishing initial conditions, an asymmetry of opposite sign is generated at early times. For small washout, this initial negative contribution is only weakly erased, leading to the counter-intuitive result that a smaller washout yields a smaller final asymmetry. At the opposite extreme, large effective neutrino masses correspond to strong washout, which erases the generated asymmetry. Between these regimes, an optimal window exists in which production is efficient while washout remains moderate, resulting in a maximal baryon asymmetry. For $M>10^5$ leptogenesis is completed well before electroweak symmetry breaking across the entire range of effective neutrino masses. Consequently, the evolution is insensitive to electroweak effects, and the corresponding curves are identical. In this regime, the curves exhibit the same functional form as the standard thermal leptogenesis efficiency factor, scaled by a large CP asymmetry \cite{Buchm_ller_2005}. We find that for $\tilde m_{max}>3.49\times 10^{-6}$ eV for $M>10^5$ GeV, the observed baryon asymmetry can be successfully reproduced. \\
Superimposed on this generic behaviour is a feature tied to the time evolution of the asymmetry itself: the baryon asymmetry undergoes an intrinsic sign change during its evolution. When electroweak symmetry breaking and the associated sphaleron freeze-out happens to occur near this sign-flip epoch, the final asymmetry becomes especially sensitive to the timing, producing the dips and structures seen for smaller $M$. As $M$ increases, the sign flip occurs at smaller $z$, so matching electroweak symmetry breaking to the sign change requires progressively smaller $\tilde m$, explaining why the features shift to smaller $\tilde m$ for larger $M$. Although this behaviour persists for even larger values of $M$, it lies outside the range shown in the figure. Since this behaviour originates from a sign change in the source term, it arises only for vanishing initial abundances. If instead thermal initial conditions, or any initial abundance larger than the thermal equilibrium value, are assumed, this feature disappears.
\subsection{Flavour Effects}
\label{sec:flavour}
\begin{figure}[H]
    \centering
    \includegraphics[width=0.7\textwidth]{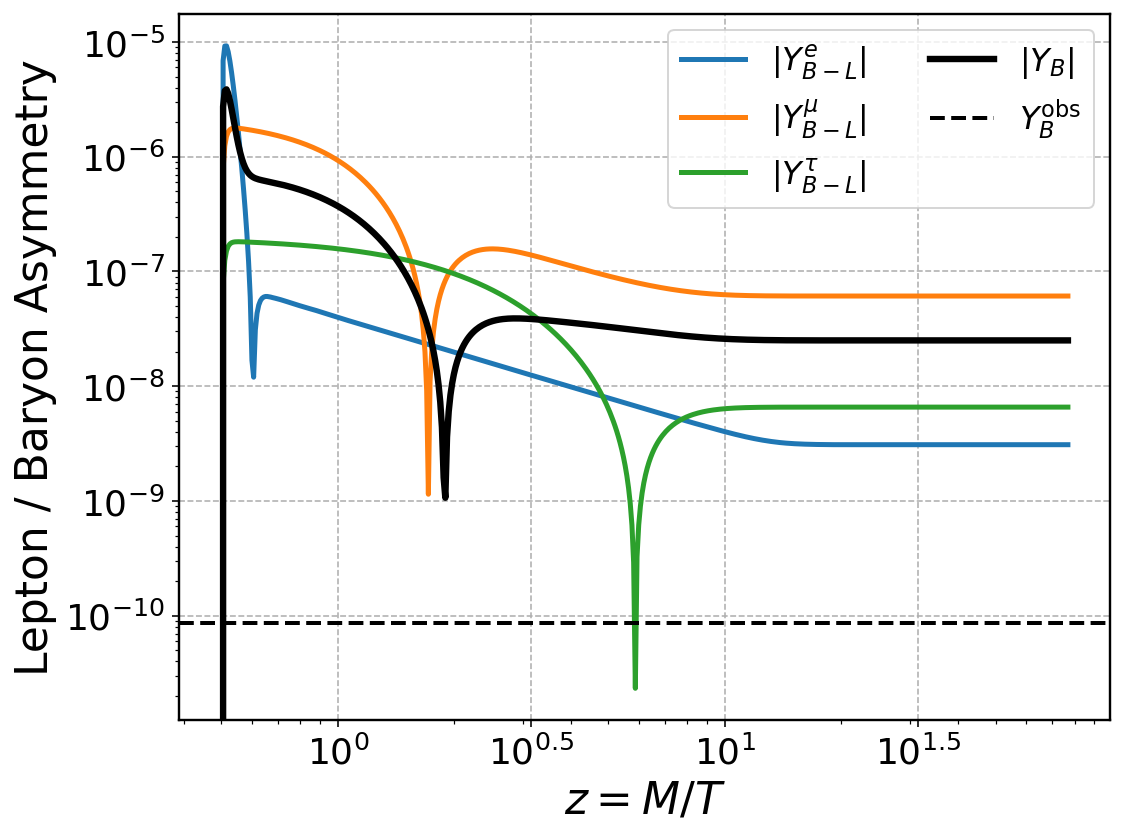}
    \caption{\it Evolution of the flavoured lepton asymmetries $|Y_{B-L}^\alpha|$ and the total baryon asymmetry $|Y_B|$ as a function of $z \equiv M/T$ for a benchmark point in fully flavoured near-resonant leptogenesis. We consider a single decaying right-handed neutrino with mass $M = 10^4~\mathrm{GeV}$ and strongly hierarchical flavoured effective neutrino masses, $(\tilde m_{1e}, \tilde m_{1\mu}, \tilde m_{1\tau}) = (1,10^{-2},10^{-4})$. The dashed horizontal line indicates the observed baryon asymmetry $Y_B^{\rm obs}$.}
    \label{fig:flavourbench}
\end{figure}

\begin{figure}[H]
    \centering
    \includegraphics[width=0.7\textwidth]{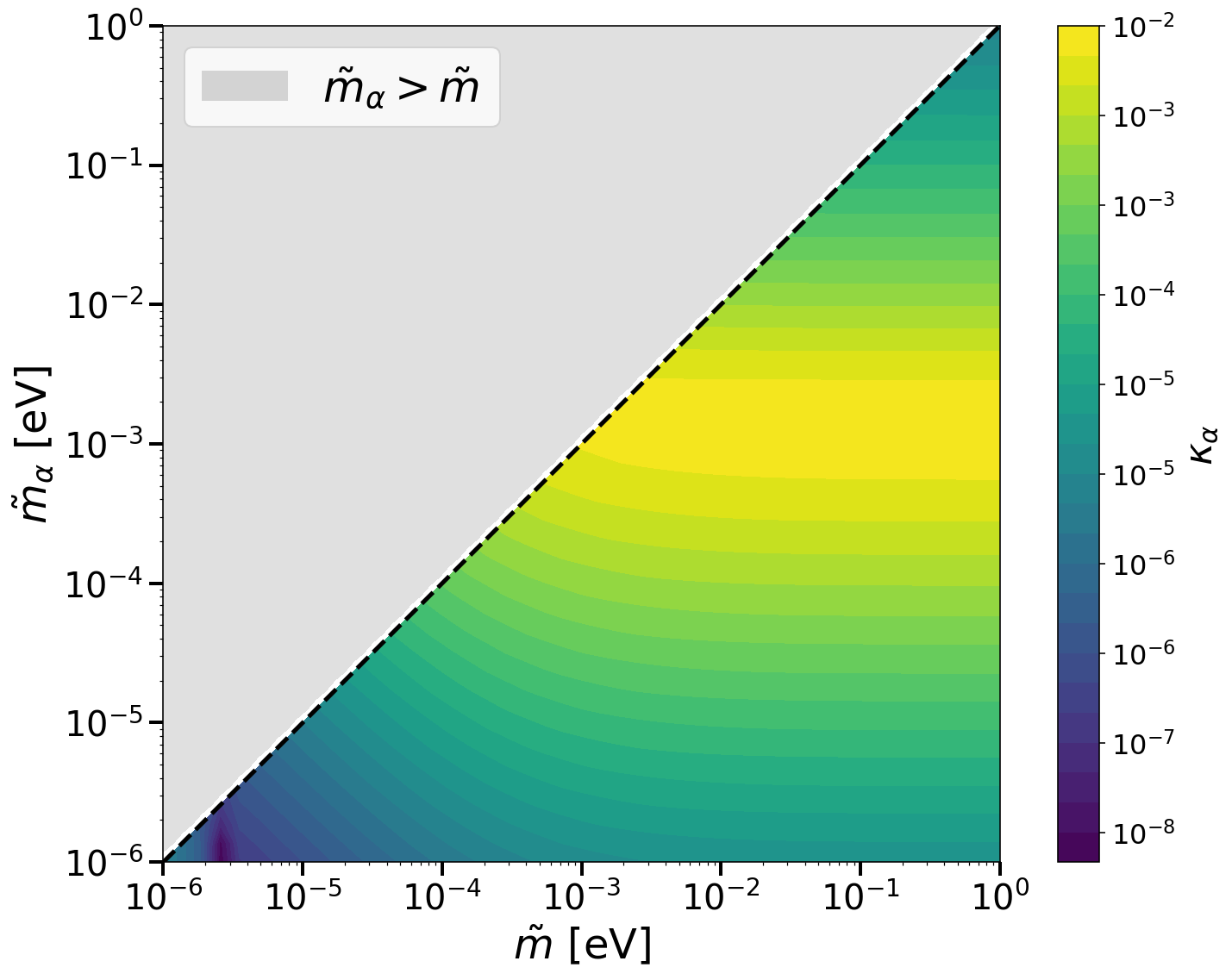}
    \caption{\it Plot of efficiency against the full effective neutrino mass, $\tilde m$, and flavoured effective neutrino mass, $\tilde m_\alpha$. The first parameter determines the production of right-handed neutrinos. The second term corresponds to the washout. In the regime where the right-handed neutrinos are produced efficiently (large $\tilde m$) the efficiency is predominantly controlled by the washout parameter. For small $\tilde m$ the production and the washout both effect the efficiency.}
    \label{fig:flavourscan}
\end{figure}

In both thermal and non-thermal cases, flavour effects play a crucial role: below temperatures of about \( 10^{12}\,\mathrm{GeV} \), the charged-lepton Yukawa interactions induce flavour decoherence, leading to distinct washout and \( CP \)-violating effects in each flavour. A consistent treatment of these flavour dynamics is therefore essential for accurately predicting the final baryon asymmetry \cite{Nardi_2006, Nardi:2006fx, Abada_2006, Antusch_2006, Blanchet_2007, DeSimone:2006nrs, Cirigliano_2010, Simone_2007, Racker_2012, Moffat_2018, baker2024hotleptogenesis, Ulysses, Ulysses2, blanchet2013leptogenesisheavyneutrinoflavours}. In the regime corresponding to $M_1 \lesssim 10^9~\mathrm{GeV}$, flavour dynamics are effectively decohered, and it is sufficient to employ the flavoured Boltzmann equations instead of the full density matrix formalism~\cite{Ulysses, Ulysses2}. The equations then take the form of the unflavoured bolztmann equation but with a flavour label and a flavour dependent washout term. We assume that the lightest right-handed neutrino has the larger effective neutrino mass. If this ordering is reversed, the labels can simply be exchanged and all expressions remain unchanged.
\begin{equation}
    \frac{dY_{B-L}^{\alpha}}{dz}=\epsilon_{1\alpha}D(z)(Y_N-Y_N^{eq})-p_{1\alpha}W(z)Y_{B-L}
\end{equation}
Here, $\alpha$ labels the lepton flavour, and $p_{1\alpha}$ denotes the probability that the heavy neutrino $N_1$ decays into a lepton of flavour $\alpha$. 
This quantity can be expressed either in terms of the Yukawa couplings or, equivalently, through the effective neutrino masses.
\begin{equation}
    p_{1\alpha}=\frac{|Y_{1\alpha}|^2}{(Y^\dag Y)_{11}}=\frac{\tilde m_{1\alpha}}{\tilde m_1}
\end{equation}
$\epsilon_{1\alpha}$ is the flavour specific CP asymmetry parameter defined in terms of flavour specific decay rates.
\begin{equation}
    \epsilon_{1\alpha}=\frac{\Gamma(N_1\rightarrow L_\alpha H)-\Gamma(N_1\rightarrow \Bar{L_\alpha} H^\dag)}{\Gamma(N_1\rightarrow L_\alpha H)+\Gamma(N_1\rightarrow \Bar{L_\alpha} H^\dag)}
\end{equation}
The flavoured CP-asymmetry in the near-resonant limit has the following limit without invoking further fine tuning on the system,
\begin{equation}
\epsilon_{1\alpha} \leq \frac{1}{200}\sqrt{p_{1\alpha}},\quad \sum_\alpha \epsilon_{1\alpha} \leq \frac{1}{200}
\end{equation}
These two bounds must hold when calculating the baryon asymmetry of the universe in this framework. These bounds along with the changed bolztmann equation to the flavoured bolztmann equation gives the framework and initial value problem to solve flavoured near-resonant leptogenesis. We provide a benchmark for a typical flavoured leptogenesis calculation shown in figure \ref{fig:flavourbench}
To quantify the efficiency of asymmetry generation in each lepton flavour, it is convenient to define a flavour-dependent efficiency factor $\kappa_\alpha$. We define
\begin{equation}
\kappa_\alpha
\;\equiv\;
\frac{79}{28}\,
\frac{|Y_{B-L}^\alpha(z_{\rm EW})|}{\epsilon_{1\alpha}Y_{\rm rel}^{\rm eq}}
\;=\;
\frac{79}{28}\,
\frac{|Y_{B}^\alpha|}{\epsilon_{1\alpha}Y_{\rm rel}^{\rm eq}} \,,
\end{equation}
where $Y_{B-L}^\alpha$ is the flavoured lepton asymmetry evaluated at the electroweak symmetry breaking temperature, $Y_{\rm rel}^{\rm eq}$ is the relativistic equilibrium abundance of right-handed neutrinos whose numerical value is $Y_{\rm rel}^{\rm eq}=3.9\times 10^{-3}$. Finally the numerical factor converts the lepton asymmetry into the baryon asymmetry. $\kappa_\alpha$ directly measures the washout efficiency in each flavour and is uniquely determined by the flavoured and total effective neutrino masses as we show in figure \ref{fig:flavourscan}.
We see that for large values of $\tilde m$ the efficiency is mainly dependent on the washout parameter $\tilde m_\alpha$, however, for smaller values corresponding the weak washout regime there is dependence on both the flavoured and total effective neutrino mass. The non-monotonic dependence of the efficiency factor $\kappa_\alpha$ on $\tilde m_\alpha$ visible in Fig.~\ref{fig:flavourscan} is a consequence of the assumed vanishing initial abundances for the right-handed neutrinos. Starting from $Y_N(z_{\rm ini})=0$, the production of right-handed neutrinos at early times implies $Y_N<Y_N^{\rm eq}$, such that the CP-violating source term initially generates a lepton asymmetry with opposite sign. If instead thermal initial conditions $Y_N(z_{\rm ini})=Y_N^{\rm eq}(z_{\rm ini})$ are assumed, no opposite sign asymmetry is generated and then $\kappa_\alpha$ decreases monotonically with $\tilde m_\alpha$, and its dependence is governed solely by washout effects.
With these efficiencies determined the final baron asymmetry is the determined by,
\begin{equation}
    Y_B=\frac{28}{79}Y_N^{\rm rel}\sum_\alpha \kappa_\alpha(\tilde m, \tilde m_\alpha)\ \epsilon_\alpha
\end{equation}
and maximising this for the strong washout regime will be able to improve the maximum baryon asymmetry produced. Without invoking flavour effects however leptogenesis is successful across all effective neutrino masses of interest. 
\section{Near-Resonant Thermal Leptogenesis during Reheating}
\label{sec:reheat}
So far, we have only considered thermal leptogenesis, where the right-handed neutrinos are produced thermally and the Universe remains radiation dominated throughout. We now turn to the case of leptogenesis during reheating where the universe begins with inflaton domination and decays into radiation ending inflation and reheating the universe. It is widely accepted that the universe underwent this phase so is a very natural stage for leptogenesis to occur. We remain focussing on thermal leptogenesis rather than non-thermal production of right-handed neutrino via inflaton decays in this work. 
\subsection{Bolztmann Equations during Reheating}
To study leptogenesis during reheating, entropy normalised abundances are no longer the most easy variable to track and neither is the time evolution variable z. This is because the temperature is no longer a one-to-one function of time, increasing during reheating before decreasing again. Instead the preferred variables are energy and number densities of the necessary quantities and the time evolution variable is the cosmological scale factor. We begin by writing the original bolztmann equations describing right-handed neutrinos, inflaton, radiation energy density and the asymmetry number density with respect to time, 
\begin{align}
    &\dot\rho_\phi + 3H\rho_\phi = -\Gamma_\phi \rho_\phi , \\
    &\dot \rho_{N} + 3H(1+w_N) \rho_{N} = -\Gamma_{N}\,(\rho_{N} - \rho_{N}^{\rm eq}) , \\
    &\dot\rho_R + 4H\rho_R = \Gamma_\phi \rho_\phi 
    + \Gamma_{N} (\rho_{N} - \rho_{N}^{\rm eq})\,\\
    &\dot n_{B-L}+3Hn_{B-L}=\frac{\Gamma_{N_i}}{\braket{E_{N}}}(\rho_{N}-\rho_{N}^{\rm eq})-W n_{B-L}
\end{align}
Where Hubble is given by the Friedmann equation and written in terms of the sum of energy densities
\begin{equation}
    H = \sqrt{\frac{8\pi}{3M_{\rm Pl}^2}\Big(\rho_\phi + \rho_{N_i} + \rho_R\Big)}\ .
\end{equation}
and the initial conditions are,
\begin{equation}
    \rho_\phi(t_I) = \tfrac{3}{8\pi} M_{\rm Pl}^2 H_I^2, 
    \qquad \rho_{N_i}(t_I) =\rho_R(t_I)=n_{B-L}=0\ .
\end{equation}
We will take $H_I=M_\phi=10^{13}$ GeV however the results of baryon asymmetry is largely independent of this. $\Gamma_\phi$ is the decay rate of the inflaton which we parametrise by the reheating temperature, $T_{RH}$ \cite{Chung:1998rq} 
\begin{equation}
\Gamma_\phi
= \sqrt{\frac{4\pi^3 g_*}{45}}\,\frac{T_{\rm RH}^2}{M_{\rm Pl}} \, .
\end{equation}
The temperature is defined by the radiation energy density
\begin{equation}
    T(t) = \Bigg( \frac{30}{\pi^2 g_*(T)}\, \rho_R(t) \Bigg)^{1/4}.
\end{equation}
The average energy per particle is calculated following maxwell-bolztmann statistics and is dependent on the temperature and mass of the right-handed neutrino
\begin{equation}
    \langle E_N \rangle = M \frac{K_1\!\left(M/T\right)}{K_2\!\left(M/T\right)} + 3T\ .
\end{equation}
The equilibrium energy density similarly calculated with mass and time dependencies, 
\begin{equation}
    \rho_N^{\rm eq}(T) \;=\; 
    \frac{1}{\pi^2} \, M^2 \, T \, K_2\!\left(\frac{M}{T}\right)
    \left[ M \frac{K_1(M/T)}{K_2(M/T)} + 3T \right] .
\end{equation}
and the equation of state parameter $w_N$ is also temperature and mass dependent,
\begin{equation}
    w_N=\frac{P_N}{\rho_N}=\frac{n_N T}{n_N \braket{E_N}}=\frac{T}{\braket{E_N}}
\end{equation}
this value interpolates between radiation $w_N=1/3$ at $T\gg M$ and matter $w_N=0$ at $T\gg M$. Finally, $W$ is the washout term given in section \ref{sec:vanilla}.
\subsubsection{Rescaling Variables}
\begin{figure}[H]
    \centering
    \includegraphics[width=\textwidth]{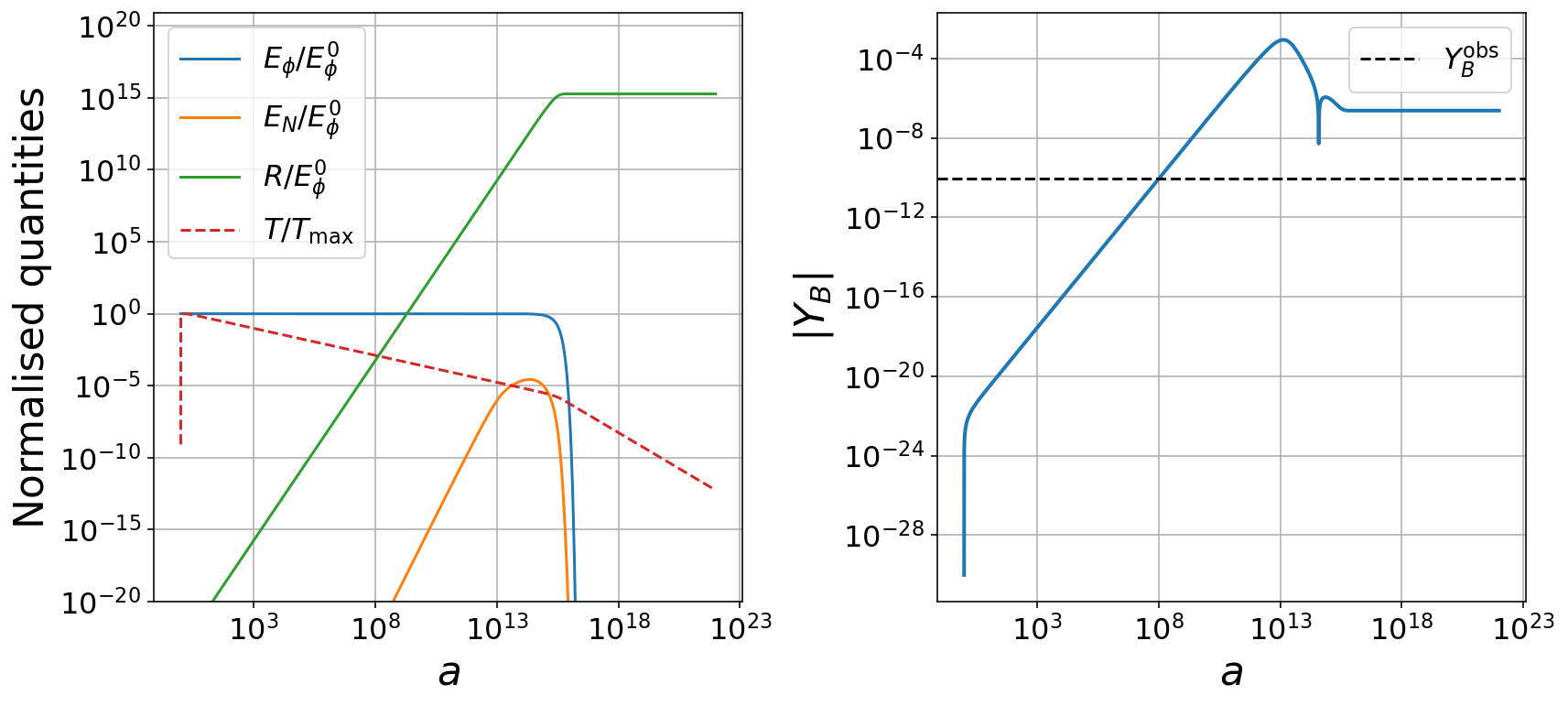}
    \caption{\it Benchmark result for near-resonant leptogenesis during reheating. The parameters are $M=10^5$ GeV, $T_{RH}=10^{4}$ GeV, $\tilde m=0.1$ eV. \textbf{Left panel:} Evolution of the normalised Boltzmann equation variables and the normalised temperature. \textbf{Right panel:} Corresponding evolution of the baryon asymmetry. Discussion of results are given in the text below.
}
    \label{fig:reheat}
\end{figure}

To make the Boltzmann equations numerically tractable it is convenient to introduce comoving variables that absorb the dilution factors due to the cosmic expansion. We define our parameters in terms of the scale factor a
\begin{equation}
    E_\phi \equiv \rho_\phi a^3, 
    \quad E_N \equiv \rho_N a^3, 
    \quad E_R \equiv \rho_R a^4 ,
    \quad N_{B-L} \equiv n_{B-L} a^3,
    \quad x \equiv \ln a ,
\end{equation}
The new variable $x=\ln a$ is the number of e-folds acting as our time evolution variable.
In terms of these comoving quantities, the Boltzmann equations take the form
\begin{align}
    \frac{dE_\phi}{dx} &= - \frac{\Gamma_\phi}{H}\, E_\phi , \\[6pt]
    \frac{dE_{N}}{dx} &=  - \frac{\Gamma_N}{H}\Big( E_{N} - E_{N}^{\rm eq} \Big) - 3 w_N\, E_{N} , \\[6pt]
    \frac{dE_R}{dx} &= \frac{a}{H}\left[ \Gamma_\phi E_\phi 
    + \Gamma_N \big( E_N - E_N^{\rm eq} \big) \right] , \\[6pt]
    \frac{dN_{B-L}}{dx} &= \frac{\epsilon\,\Gamma_N}{H}
    \frac{E_N - E_N^{\rm eq}}{\langle E_N \rangle} -\frac{W}{H}N_{B-L}\, .
\end{align}
The Hubble friction terms $3H$ and $4H$ have disappeared however with the $w_N$ dilution term still remaining, these
equations are still much better behaved numerically. The Hubble rate is determined self-consistently from the total energy density, which we can write in terms of comoving variables as
\begin{equation}
    H(x) = \sqrt{\frac{8\pi}{3M_{\rm Pl}^2}\left( 
    \frac{E_\phi}{a^3} + \frac{E_N}{a^3} + \frac{E_R}{a^4} \right)} ,
\end{equation}
with $a = e^x$. The radiation bath is assumed to thermalises very efficiently, and its temperature is determined by the radiation energy density 
\begin{equation}
    T(x) = \left( \frac{30}{\pi^2 g_*(T)}\, \frac{R}{a^4} \right)^{1/4}.
\end{equation}
The remaining terms are then functions of the temperature and right-handed neutrino mass. The initial conditions come from inflation giving the energy budget entirely to the inflaton with vanishing initial conditions from 
\begin{equation}
    E_\phi(x_I) = \frac{3}{8\pi} M_{\rm Pl}^2 H_I^2 \, a_I^3, 
    \qquad E_N(x_I) = E_R(x_I)=N_{B-L}=0\ .
\end{equation}
We take $a_I=1$ as nothing physical depends on this choice. Finally, the observed baryon asymmetry is obtained from the left-handed asymmetry via the sphaleron conversion factor
\begin{equation}
    Y_B = \frac{28}{79}\,\frac{N_{B-L}/a^3}{s}.
\end{equation}
where $s$ is the entropy density. This is the initial value problem for near-resonant leptogenesis during reheating. To gain intuition for this behaviour, we present a benchmark example showing the evolution of the relevant quantities shown in figure \ref{fig:reheat}.
The inflaton initially dominates the energy density and remains approximately constant until its decay rate becomes comparable to the Hubble expansion rate, after which it decays into radiation completing reheating. The radiation energy density grows logarithmically and approaches a constant comoving value after the inflaton decays have completed. The right-handed neutrino abundance increases from inverse decays once the temperature decreases enough for the Lorentz thermal averaging to become non-lethal on the decay rate. The right-handed neutrinos then become non-relativistic and decay to a negligible abundance. The temperature initially vanishes and rises rapidly as radiation is produced from inflaton decay, reaching a maximum value. It subsequently decreases as the Universe expands; however, this cooling is partially compensated by the continued growth of the comoving radiation density, leading to a comparatively moderate decline. Once the comoving radiation density becomes constant, the temperature redshifts with the expansion and decreases at a more rapid rate. The baryon asymmetry is initially generated from inverse decays, when the decays begin the asymmetry undergoes a change of sign. Once the right-handed neutrino and inflaton decays have completed the baryon asymmetry is frozen in place.
\subsection{Bound on Reheating Temperature}
\begin{figure}[H]
    \centering
    \includegraphics[width=\textwidth]{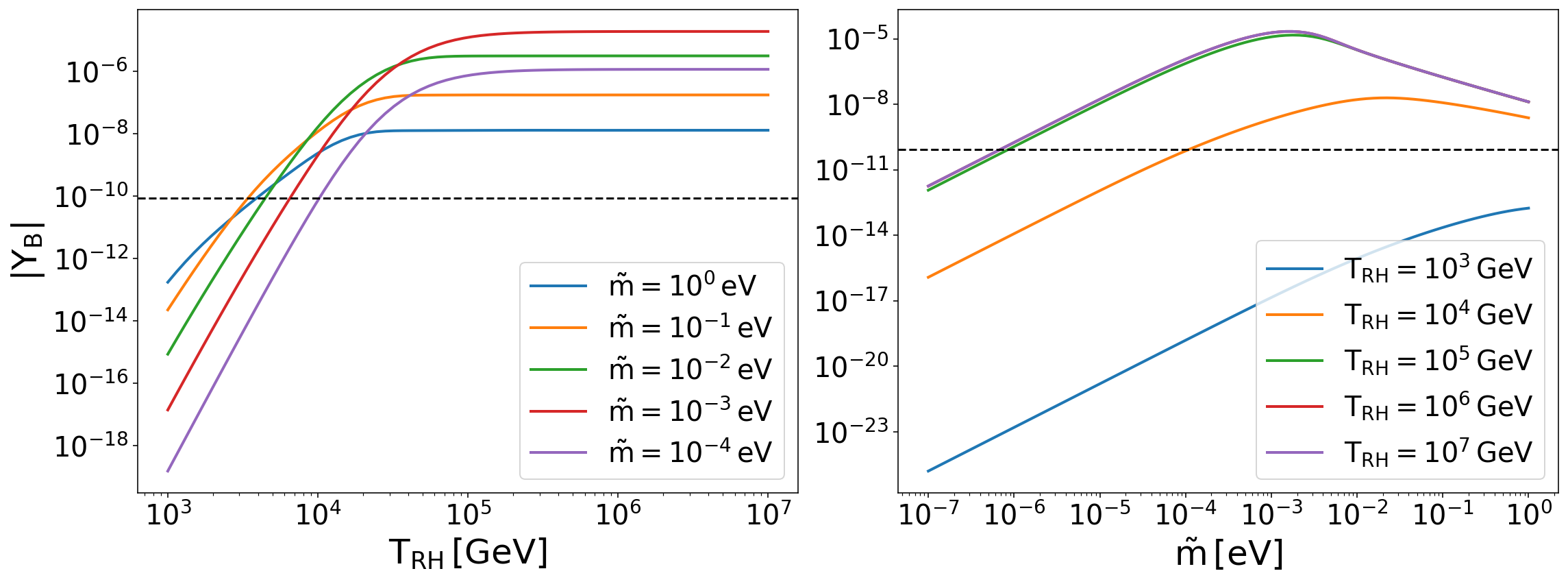}
    \caption{\it Scan over reheating temperatures and effective neutrino masses for near-resonant leptogenesis during reheating for $M=10^5$ GeV. \textbf{Left panel:} shows the baryon asymmetry dependence on the reheating temperature for various $\tilde m$ values demonstrating leptogenesis favour larger reheating temperatures. We find that the baryon asymmetry monotonically increases with reheating temperature until the reheating temperature exceeds the mass of the right-handed neutrino, then we recover the results for section \ref{sec:vanilla}. \textbf{Right panel:} shows the dependence on effective neutrino mass for fixed reheating temperatures. For large reheating temperatures we recover the dependence on effective neutrino mass in figure \ref{fig:thermalscan3} whereas when the reheating temperature drops the baryon asymmetry drops and dependence on effective neutrino mass changes. Further discussions are given in the text below. In both plots the dashed black line denotes the observed asymmetry. 
}
    \label{fig:reheatscan}
\end{figure}

\begin{figure}[H]
    \centering
    \includegraphics[width=0.49\textwidth]{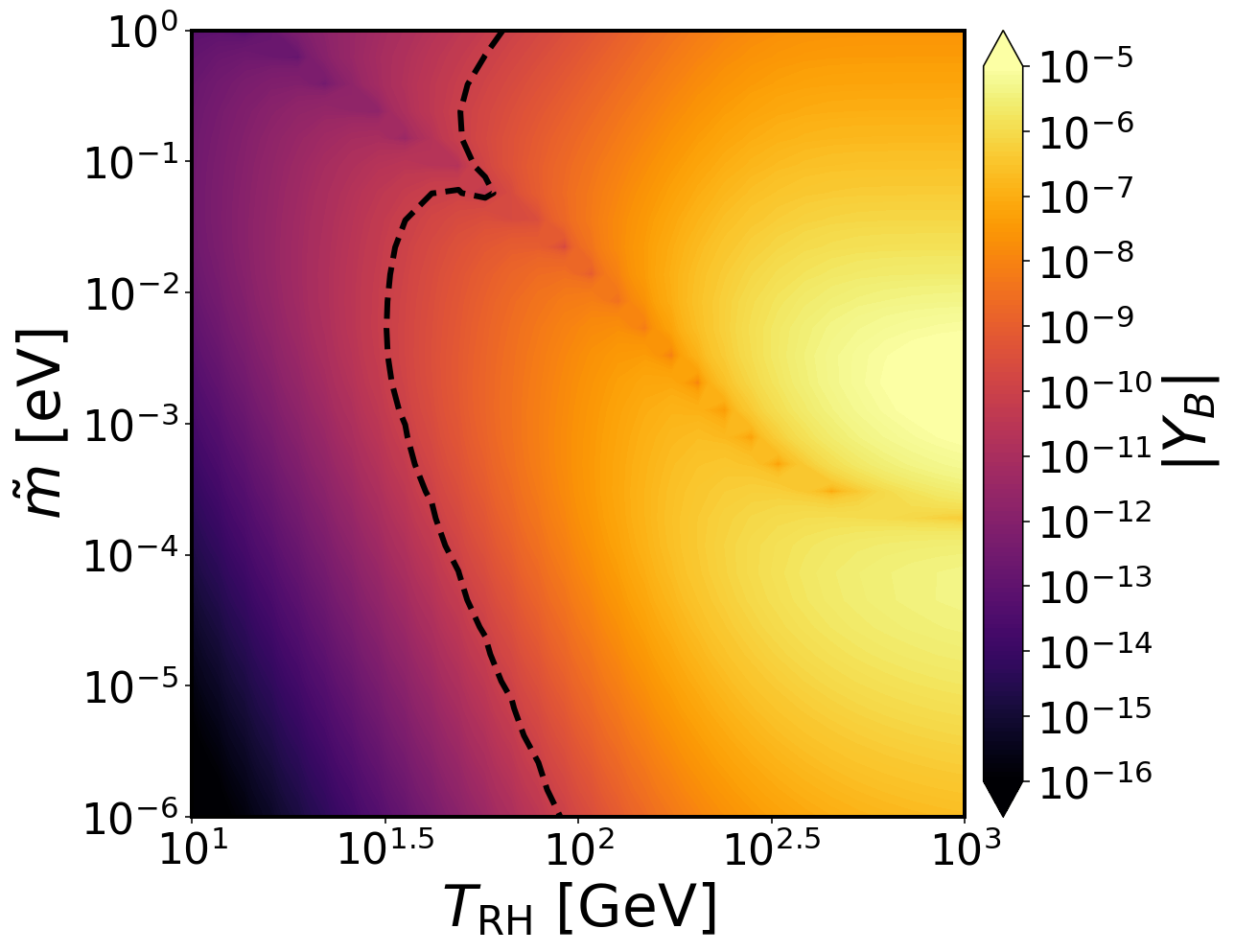}
    \includegraphics[width=0.49\textwidth]{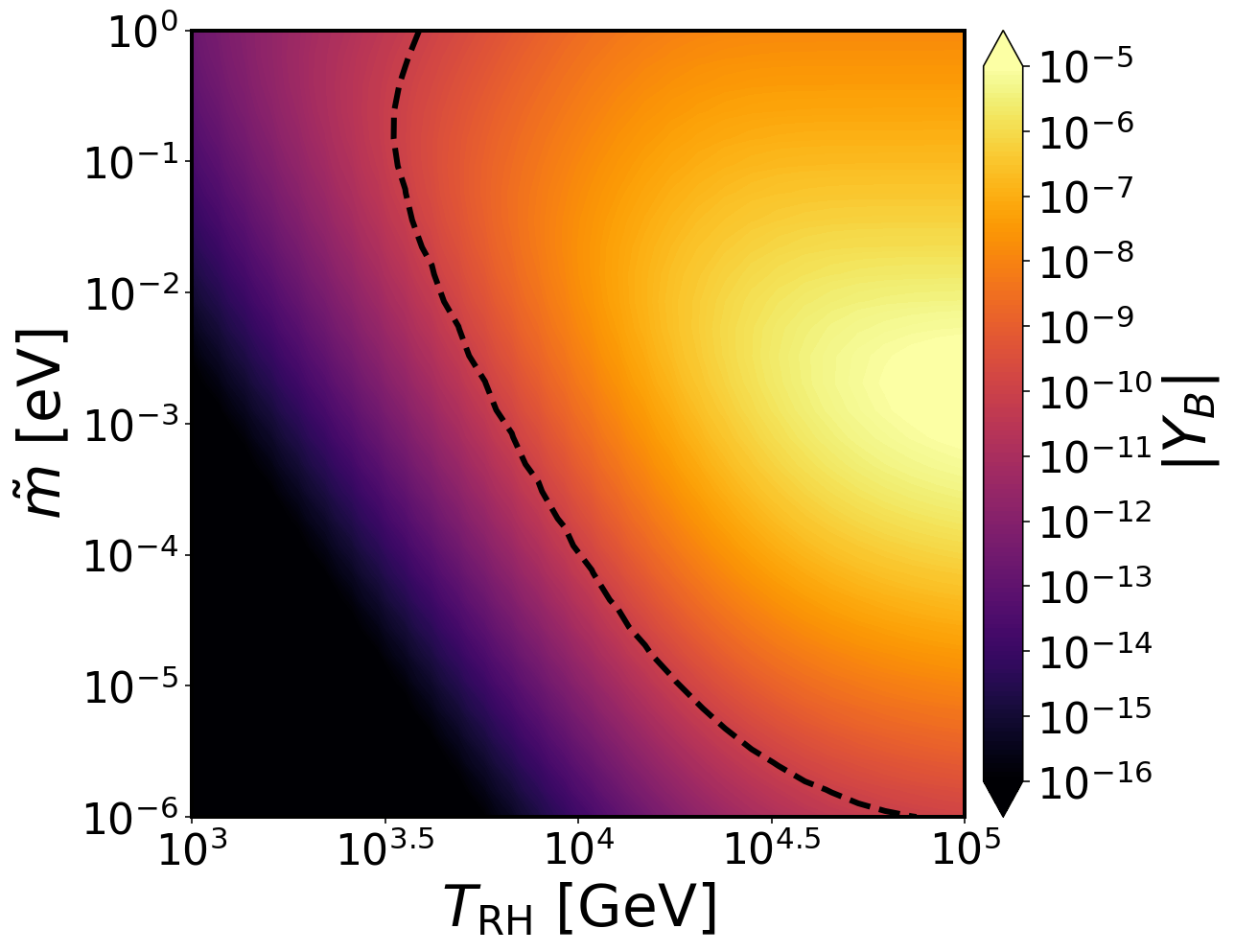}
    \caption{\it Scan for near-resonant leptogenesis during reheating scanning over reheating temperatures, $T_{RH}$, and effective neutrino masses, $\tilde m$. \textbf{Left panel:} $M=10^3$ GeV we find the minimum reheating temperature is $35.6$ GeV below the electroweak temperature however leptogenesis can still be successful as the reheating temperature is not the maximum temperature and a large enough asymmetry has been generated before electroweak symmetry breaking. \textbf{Right panel:} $M=10^5$ GeV showing the full scan slices of which were shown in figure \ref{fig:reheatscan}. For this mass the lowest reheating temperature is $3.56$ TeV however does not occur at the same effective neutrino mass. The black dashed line denotes the observed asymmetry, all points to the right of the line can yield successful near-resonant leptogenesis.  
}
    \label{fig:reheatscan2}
\end{figure}

The purpose of this subsection is to investigate the effect the reheating temperature has on near-resonant leptogenesis. To investigate this in Fig.~\ref{fig:reheatscan} we present parameter scans in which either the reheating temperature or the effective neutrino mass is held fixed while varying the other for a fixed Right-Handed neutrino mass.
The plot on the left shows that the baryon asymmetry increases monotonically with reheating temperature. For reheating temperatures greater than the mass of the decaying right-handed neutrino leptogenesis occurs during radiation domination as leptogenesis largely occurs for $T\lesssim M$ so we recover results for thermal leptogenesis in section \ref{sec:vanilla}. For smaller values of the reheating temperature there are two factors leading to a smaller asymmetry, right-handed neutrino production and entropy dilution. When the reheating temperature falls below the right-handed neutrino mass scale, the lepton asymmetry is generated prior to the main entropy release from inflaton decays and is therefore diluted, resulting in a reduced final asymmetry. Moreover, right-handed neutrino production becomes increasingly inefficient at low reheating temperatures, although this effect is partially alleviated for larger effective neutrino masses. This behaviour is also visible in the right panel. For large reheating temperatures, the dependence on the effective neutrino mass matches that of Fig.~\ref{fig:thermalscan3}. By contrast, at smaller reheating temperatures the asymmetry is strongly suppressed and its dependence on the effective neutrino mass changes, favouring larger values of $\tilde m$.\\
Recalling that the reheating temperature is not the maximum temperature of the universes history, the efficiency of near-resonant leptogenesis therefore gives the possibility of a reheating temperature less than the electroweak temperature as baryogenesis occurs before electroweak symmetry breaking at that point the baryon number density is frozen and entropy dilution dilutes the observed baryon asymmetry abundance. To calculate this we scan over variables of interest and for each benchmark we calculate the baryon asymmetry at electroweak symmetry breaking and input entropy dilution by hand.   
\begin{equation}
Y_B \;\equiv\; \frac{n_B}{s}
\;=\; \frac{28}{79}\,\frac{N_{B-L}(x_{\rm EW})}{S(x_{\rm RH})}
\;=\; \frac{28}{79}\,\frac{N_{B-L}(x_{\rm EW})}{a^3(x_{\rm RH})\,s\!\left(T_{\rm RH}\right)} \, ,
\end{equation}
We know give full scans over the parameter space for multiple right-handed neutrino masses shown in figure \ref{fig:reheatscan2} showing that reheating temperatures less than the electroweak symmetry breaking temperature can give successful leptogenesis.
We find similar results and tendencies described from the last plot with the baryon asymmetry increasing with reheating temperature. The additional feature of interest is concerning the $M=1$ TeV plot there is the dip where the electroweak symmetry breaking occurs during the sign flip and is dependent on both the reheating temperature and the effective neutrino mass. This feature was discussed at length in section \ref{sec:vanilla}. For larger values of reheating temperature we will recover the results from \ref{sec:vanilla} and the dependence of the feature will be solely the effective neutrino mass. A full scan was completed and it was found that the reheating temperature can be as low as $T_{RH}\approx 10$ GeV for Masses around the electroweak scale. Generally however for Masses at the TeV scale and above the minimum reheating temperature is roughly twenty eight times less than the Mass. For $M=10^4$ GeV the value the minimum reheating temperature is $T_{RH}=386$ GeV so the value is not a fixed proportionality to Mass and in no case are the effective neutrino mass leading to this minimum the same for Masses. However, generally the minimum reheating temperature occurs at larger effective neutrino masses for smaller right-handed neutrino masses.   
\subsection{Flavour Effects}
\begin{figure}[H]
    \centering
    \includegraphics[width=0.7\textwidth]{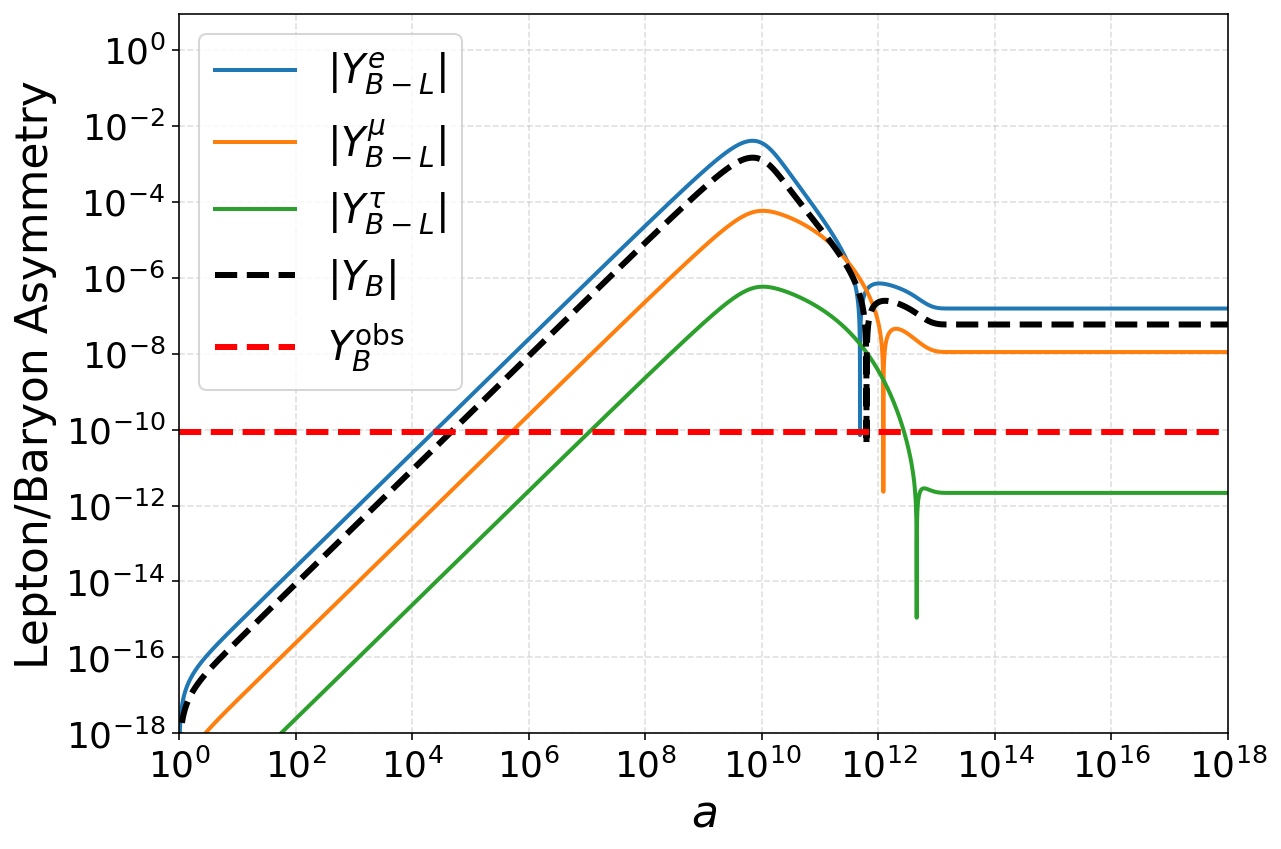}
    \caption{\it Benchmark evolution of the flavoured lepton asymmetries and the resulting baryon asymmetry for $M=10^7$ GeV, $T_{RH}=10^6$ GeV and $\tilde m_\alpha=(1,10^{-2}, 10^-4)$ eV. The dashed red line corresponds to the observed baryon asymmetry.}
    \label{fig:reheatflavour}
\end{figure}

Flavour effects during reheating can be incorporated in direct analogy with their treatment in thermal near-resonant leptogenesis. Since flavour dynamics have already been discussed extensively earlier in this work, we restrict ourselves here to outlining the structure of the flavour-resolved equations relevant during reheating. For $M \lesssim 10^9\,\mathrm{GeV}$, the use of flavour-dependent Boltzmann equations provides an accurate description of the evolution of the lepton asymmetry. In terms of the comoving variables introduced above, the equations generalise to
\begin{equation}
    \frac{dN_{B-L}^\alpha}{dx} =
    \frac{\epsilon^\alpha\,\Gamma_N}{H}
    \frac{E_N - E_N^{\rm eq}}{\langle E_N \rangle}
    - \frac{W^\alpha}{H} N_{B-L}^\alpha \, ,
\end{equation}
where $\alpha=e,\mu,\tau$ labels lepton flavour and $N_{B-L}=\sum_\alpha N_{B-L}^\alpha$. All flavour dependence enters through the CP asymmetries $\epsilon^\alpha$ and the washout rates $W^\alpha$, defined as in the radiation-dominated case but evaluated on the reheating background. The remaining bolztmann equations and initial conditions remain unchanged. We provide a benchmark of the lepton and baryon asymmetries in figure \ref{fig:reheatflavour},
The asymmetries initially grow slowly due to Lorentz thermal averaging, before inverse decays become efficient. Once the right-handed neutrinos become non-relativistic, decays dominate and induce a change in sign. The individual lepton flavour asymmetries differ as each experiences a distinct washout strength. So long as the washout is not negligible these effects should be taken into account for an accurate prediction of the baryon asymmetry. Since flavour dynamics during reheating have not yet been fully explored even in the case of hierarchical right-handed neutrino masses, we leave such an investigation to future work currently in progress.

\section{Conclusions}
\label{sec:conclusion}
In this work we have analysed leptogenesis in the quasi-degenerate but explicitly non-resonant regime, with the aim of clarifying how far decay-driven leptogenesis can be pushed in mass scale and reheating temperature without invoking resonant enhancement. By expanding the CP asymmetry close to degeneracy and imposing a conservative non-resonance condition, $\Delta M > a\,\Gamma_i$ gives a universal upper bound on the CP asymmetry, $\epsilon < 1/2a$, which is independent of the effective neutrino masses and the right-handed neutrino mass. This bound provides a robust, model-independent benchmark for the maximal CP violation available in the non-resonant quasi-degenerate regime. 
While the CP-asymmetry bound was previously derived in an extended $U(1)_{B-L}$ framework \cite{datta2025gravitationalwavespectralshapes,ghoshal2025cosmicstringsgravitationalwave}, the present analysis shows how it can be consistently implemented within the minimal type-I seesaw and tested across radiation domination, with flavour effects included, and in reheating cosmologies. Most significantly, we find that this controlled quasi-degenerate, explicitly non-resonant regime lowers the viable leptogenesis scale to the electroweak scale. In particular, successful leptogenesis from right-handed neutrino decays is achievable for $M_1 \gtrsim 100~\mathrm{GeV}$, with the bound remaining largely independent of the effective neutrino masses. Parameter scans over both effective neutrino masses and heavy-neutrino mass scales confirm the robustness of this result, as shown in Figs.~\ref{fig:thermalscan1} and \ref{fig:thermalscan3}.\\
We have also clarified the role of the non-resonance parameter $a$ entering the condition $\Delta M > a\,\Gamma_i$, which leads to the general bound $|\epsilon| < 1/(2a)$. In this work we fix $a=100$ as a conservative benchmark choice. The generated baryon asymmetry scales proportionally to $1/a$, and varying $a$ therefore rescales the overall normalisation of all baryon asymmetry curves. However, throughout the viable parameter space the produced asymmetry is comfortably larger than the observed value before imposing this normalisation. As a result, moderate variations of $a$ do not determine whether leptogenesis is successful, nor do they alter the conclusion that decay-driven leptogenesis can operate down to the electroweak scale within the non-resonant regime.\\
We then investigated near-resonant thermal leptogenesis during reheating. We provided benchmarks and scans over the parameter space showing that successful leptogenesis can occur for reheating temperatures up to twenty eight times less than the right-handed neutrino mass in figure \ref{fig:reheatscan2}. We were able to utilise the fact that the reheating temperature is not the maximum temperature to have reheating temperatures below the electroweak symmetry breaking temperature whilst still having successful leptogenesis.  We found successful leptogenesis can occur for reheating temperatures as low as around $T_{RH}\simeq 10$ GeV without any non-thermal production mechanisms. Finally we provided the framework and initial value problem for flavoured near-resonant thermal leptogenesis during reheating.\\
Taken together, our results demonstrate that quasi-degenerate, non-resonant leptogenesis can operate successfully at mass scales down to those typically associated with ARS leptogenesis, while remaining compatible with a wide range of reheating temperatures spanning more than an order of magnitude below and above the mass scale. This substantially broadens the class of cosmological scenarios consistent with non-resonant, decay-driven leptogenesis. More generally, near-resonant thermal leptogenesis provides a theoretically stable alternative to resonant leptogenesis for lowering the leptogenesis scale from $10^9$ GeV down to the electroweak scale, without relying on a disputed regulator inherent to resonant leptogenesis. 
\section*{Acknowledgements}
I thank Graham White for helpful comments on the manuscript and for encouragement throughout the project. I also wish to thank Pasquale Di Bari for discussions on leptogenesis and flavour effects. I acknowledge the STFC Consolidated Grant ST/T000583/1 and I thank the University of Southampton School of Physics and Astronomy for the support of a Mayflower PhD scholarship.
\bibliographystyle{unsrt}
\bibliography{main}
\end{document}